\documentclass[prb,twocolumn,showpacs]{revtex4}
\usepackage{epsfig,latexsym,amssymb,amsfonts,amsmath}

\begin{document}
\title{The ordered phase of  the one-dimensional Ising spin glass with
long-range  interactions} \author{M.~A. Moore}  \affiliation{School of
Physics and  Astronomy, University of Manchester,  Manchester M13 9PL,
UK}

%\date{\today}

\begin{abstract}
The  one-dimensional  long-range  Ising  spin  glass  provides  useful
insights into  the properties of finite dimensional  spin glasses with
short-range  interactions.  The  defect  energy renormalization  group
equations  derived for  it by  Kotliar, Anderson  and Stein  have been
re-examined and  a new fixed point  has been found.   The fixed points
which have previously been studied  are found to be inappropriate.  It
is shown that the  renormalization group equations themselves directly
imply  that  that  the  spin  glass phase  is  replica  symmetric,  in
agreement  with droplet  model expectations,  when the  range exponent
$\sigma$ lies between the upper and lower critical values.

\end{abstract}

 \maketitle
\section{Introduction}
\label{sec:intro}
In recent  years there have  been many simulations  of one-dimensional
Ising      spin     glasses      with      long-range     interactions\cite{Leuzzi:99,KY1,KY2,Leuzzi:08,Hart08,Larson}. The Hamiltonian
of these studies is usually a variant of the following:
\begin{equation}
H=-\sum_{<ij>}J_{ij}S_iS_j,
\label{Ham}
\end{equation}
where the  Ising spin $S_i$  takes on the  values $\pm 1$, the  sum is
 over  all  pairs  $<ij>$,  and  $i$  and $j$  are  positions  on  the
 one-dimensional lattice. The interaction
\begin{equation}
J_{ij}=J\frac{\epsilon_{ij}}{|i-j|^{\sigma}},
\label{sigma}
\end{equation}
where  the $\epsilon_{ij}$  are  independent random  variables with  a
Gaussian distribution of zero mean  and unit variance.  This model was
introduced by  Kotliar, Anderson and Stein \cite{KAS},  to be referred
to as KAS, who showed that  for $\sigma< 2/3$ the model has mean-field
critical exponents, and non-mean  field exponents for $2/3 <\sigma <
1$.   When   $\sigma > 1$,  there  is  no   finite  temperature  phase
transition.  Thus the  $\sigma$-interval, $2/3  <\sigma <  1$,  is the
analogue for  short-range spin glasses of the  dimension range between
the  upper  critical  dimension   ($d_u=6$)  and  the  lower  critical
dimension.  A rough correspondence between  a value of $\sigma$ in the
long-range one-dimensional  model and  the space dimension  $d$ in  the short-range
model has been suggested \cite{Larson, Larsonpspin}:
\begin{equation}
d=\frac{2-\eta(d)}{2\sigma-1},
\label{map}
\end{equation}
where $\eta(d)$ is the critical exponent of the short-range model.
 
One  motivation   for  studies  of  the   one-dimensional  model  with
long-range interactions is that by just changing the value of $\sigma$
one  can   explore  both  systems   corresponding  to  high   and  low
dimensionality.  Thus  study of this  long-range one-dimensional model
can cast light on the  long-standing expectation that the whole nature
of  the  short-range  spin  glass  state  changes  at  six  dimensions
\cite{Moore1}. It is impossible to imagine doing a good simulation for
a ten-dimensional  system, but simulations at  the corresponding value
of $\sigma$, which is $0.6$, are no harder than those at, say, $\sigma
=0.75$, which corresponds to a dimension less than $6$.

 Above six dimensions the ordered phase is expected to be described by
Parisi's replica symmetry  breaking procedure \cite{Parisi}. There the standard procedure of performing the loop expansion around the mean-field solution should suffice. Below six
dimensions it  has been argued that the spin glass phase is  replica symmetric and
described by  the droplet picture  \cite{McM, BM, FH}. Thus  above six
dimensions,  there  is  expected  to  be a  de  Almeida-Thouless  (AT)
transition  in  a magnetic  field,  but  none  is expected  below  six
dimensions. In the one-dimensional model, strong simulational evidence
was  found for  an AT  transition for  $\sigma =0.6$,  but  at $\sigma
=0.75$    there     was    no    sign    of     an    AT    transition
\cite{Larson}. Unfortunately,  as so often with  numerical studies, it
is possible  to find another  way of analysing  the data to  reach the
opposite conclusion \cite{Leuzzi2009}. What  is needed therefore is an
analytical approach which can settle the long-running debate between the
simulators.

For these  long-range one-dimensional  models  KAS found
  an analytic treatment which should be quantitative near the
lower  critical value of  $\sigma$, which is $1$.   (KAS also  studied
critical behavior  in the  vicinity of the  ``upper critical  value of
$\sigma$'',  (which is  $2/3$) by  means of  conventional perturbative
expansions in $(\sigma - 2/3)$).   In this paper we shall follow their
second  renormalization  group   (RG)  procedure,  which  involves  an
extension  of equations  given  by Cardy  \cite{Cardy},  to obtain  an
expansion  about the  ``lower critical  value''. Expansions  about the
lower  critical dimension have  never been  done for  short-range spin
glasses,  probably because  not even  the precise  value of  the lower
critical   dimension  is   known   for  them!    For  the   long-range
one-dimensional  model,  KAS  derived  the  RG  flow  equations  which
describe the flows  of kink (or  defect or droplet)  couplings and
their associated  fugacities not only near  the transition temperature
$T_c$,  but  also  throughout  the low-temperature  phase.   Thus  the
solution of these equations should  allow us to determine not only the
values of  the critical exponents but  also the nature  of the ordered
phase.  We shall employ a technique used previously\cite{Moore2} to show 
that  the spin glass phase has replica symmetry
when $2/3<\sigma  <1$. This is a direct consequence of the 
 KAS equations themselves  and does not require us to find 
an explicit solution of them, which is fortunate as we are unable
 to provide one.

The RG  equations of  KAS involve the  replica procedure and  are very
difficult to  solve. In Section  \ref{formalism} it is shown  that the
fixed point  solution found by  KAS themselves is not  the appropriate
fixed point -- it actually is relevant to a multicritical point -- and
does not describe the  behavior across the paramagnet-spin glass phase
boundary.  Khurana \cite{AK} introduced important ideas which simplify
the task  of solving  the KAS RG  equations. The calculations  of this
paper in Sections \ref{formalism}  and \ref{flows} are an extension of
them. The actual fixed point found by Khurana is not the
correct one when  the number of replicas $n <2$;  in the replica trick
we have to take the limit $n \rightarrow 0$.  In Section \ref{flows} a
solution of  the KAS RG equations  is obtained which is  valid for the
limit $\sigma \rightarrow 1$ and when $T \le T_c$.

 In  Section \ref{sec:field} the  RG flow equations are solved when  a small
random magnetic field  is included in the Hamiltonian,  and  the
magnetic critical  exponent is obtained. In Section \ref{sec:PQ} it  is shown that
the spin glass phase has replica symmetry for $2/3 <\sigma <1$.  This is the main result
of the paper.

\section{Renormalization Group Equations}
\label{formalism}
We begin with a brief outline of the procedure of KAS by which they
obtained  an expansion  around the lower critical value of $\sigma  (=1)$. They used
the replica  method to write  the average over the  $\epsilon_{ij}$ of
the $n$th power of the partition function as
\begin{multline}
\overline{Z^n}=\mbox{Tr}\exp[-\beta  H_n]\\={\rm  Tr}_{S^{\alpha}_{i}}
\exp\left(\frac{1}{2}\beta^2J^2\sum_{i<j}\sum_{\alpha,\beta}S_i^{\alpha}S_i^{\beta}S_j^{\alpha}S_j^{\beta}/
|i-j|^{2\sigma}\right).
\label{RP}
\end{multline}
The free energy
\begin{equation}
F=-(\beta n)^{-1}(\overline{Z^n}-1),
\label{feq}
\end{equation}
as $n \rightarrow  0$. At each site $i$ there  are the replicated spin
 variables $S_i^{\alpha}$, where $\alpha =1,2, \cdots,n$. The possible
 values of  $S_i^{\alpha}$ lie at  the vertices of  an $n$-dimensional
 hypercube.       KAS       introduced      $n$-component      vectors
 $\boldsymbol{\sigma}_{a}$,  $a =1,\cdots,2^n$  to describe  the $2^n$
 different       values        of       $\{S_i^{\alpha}\}$,       i.e.
 $\boldsymbol{\sigma}_1=(1,1,\cdots,1,1)$,
 $\boldsymbol{\sigma}_2=(1,1,\cdots,1,-1)$,
 $\boldsymbol{\sigma}_3=(1,1,\cdots,-1,-1)$  etc..   In the  following,
 replica indices which  run from $1, 2, \cdots, n$  will be denoted by
 Greek symbols  like $\alpha$  or $\beta$. Labels  which run  over the
 $2^{n}$  spin states  will be  denoted by  Roman labels  like  $a$ or
 $b$. The interaction  energy between a spin in state  $a$ at site $i$
 and a spin in state $b$ at site $j$ is
\begin{equation}
\frac{K_{a
b}}{|i-j|^{2\sigma}}=\frac{1}{2}\beta^2J^2\frac{[(\boldsymbol{\sigma}_{a}\cdot\boldsymbol{\sigma}_{b})^2-n^2]}{|i-j|^{2\sigma}},
\label{Kdef}
\end{equation}
where a  constant, $n^2$, has  been subtracted to ensure  that $K_{aa}
\equiv 0$.  KAS  wrote the replicated Hamiltonian in  terms of kink or
defect variables; a kink  of type $ab$ is at site $i$  if that site is
in  state  $\boldsymbol{\sigma}_{a}$  and   site  $i+1$  is  in  state
$\boldsymbol{\sigma}_{b}$.   For  each  kink  type $ab$  there  is  an
associated fugacity $Y_{ab}$, and  $Y_{aa} \equiv 0$.  KAS showed that
a  change  in  the  lattice   spacing  $a  \rightarrow  ae^l$  can  be
compensated  by a  change in  the  kink fugacities  $Y_{ab}$ and  kink
couplings  $K_{ab}$   to  leave  invariant   the  replicated  averaged
partition  function of  Eq.  (\ref{RP}).   This leads  to the  RG flow
equations
\begin{equation}
\frac{dY_{ab}}{dl}=Y_{ab}\left(1+\frac{2K_{ab}}{2\sigma-1}\right)+\sum_{c\neq
a,b}Y_{ac}Y_{cb},
\label{RGY}
\end{equation}
\begin{eqnarray}
\frac{dK_{ab}}{dl}&=&2(1-\sigma)K_{a
b}-\sum_{c}Y^2_{ac}(K_{ab}-K_{bc}+K_{ac})         \nonumber         \\
&-&\sum_{c}Y^2_{bc} (K_{ab}-K_{ac}+K_{bc}).
\label{RGK}
\end{eqnarray}
These equations are exact  for small fugacities. When
 $\sigma$ is close to $1$, the effect of the fugacities does indeed turn out to be small. As $\sigma$ decreases 
towards the ``upper critical value'' $2/3$,  higher terms in $Y_{ab}$ will be needed.
But the expectation is that the model for $\sigma$ just below $1$ is qualitatively 
similar to a model with $\sigma$ just above $2/3$. Thus if one can establish that the 
spin glass phase is replica symmetric using Eqs. (\ref{RGY}) and (\ref{RGK}), (as we can), then this behavior
should hold up to the upper critical value $2/3$.
 
  One normally proceeds
by  starting  off the  RG  flows  from the  bare  values  of the  kink
couplings $K_{ab}$ and the corresponding bare fugacities $Y_{ab}$. The
initial value of the kink couplings $K_{ab}$
\begin{equation}
K_{ab}(l=0)=\frac{\beta^2J^2}{2}k(p),
\label{bare}
\end{equation}
where using Eq. (\ref{Kdef})
\begin{equation} 
k(p)=[(n-2p)^2-n^2]=-4p(n-p).
\label{defkp}
\end{equation}
Here $p$  is the  number of  replicas in which  spins at  two adjacent
sites, in  state $a$ at  $i$ and $b$  at $i+1$, are  anti-parallel and
takes the values $p=1,2, \cdots,n$.

One can  use symmetry considerations  to simplify the  $2^n\times 2^n$
matrices $K_{ab}$ and $Y_{ab}$.  Re-numbering the replicas should make
no   physical  difference.    Formally  permutations   $\pi$   of  the
permutation  group   ${\cal  S}_n$  acting  on   the  replica  indices
${1,\cdots,n}$ must result in
\begin{equation}
K(\pi(\boldsymbol{\sigma)},\pi(\boldsymbol{\sigma}^{\prime}))=K(\boldsymbol{\sigma},
\boldsymbol{\sigma}^{\prime})  \hspace{0.5cm}  \mbox{for every}\:  \pi
\in {\cal S}_n.
\label{perm}
\end{equation}
   As  the  bare  couplings  in  Eq.  (\ref{Kdef})  are  of  the  form
 $K(\boldsymbol{\sigma},\boldsymbol{\sigma}^{\prime})=K(\boldsymbol{\sigma}\cdot
 \boldsymbol{\sigma}^{\prime})$  we shall  suppose  that this  feature
 remains true of  the renormalized kink couplings; it  is preserved by
 the RG flow  equations of Eqs.  (\ref{RGY}) and  (\ref{RGK}).  An example
 might make  clearer the  simplification which these  symmetries allow.
 Consider the case of $n=2$. Then the matrix $K_{ab}$ is a $4\times 4$
 matrix.  A state  like $a$ corresponds to one of $00,01,10,11$  where a down
 spin  is represented by  $0$ rather  than $-1$  to ease  layout.  The
 elements of  $K_{ab}$ are:\\ $K_{00,00}=0,  K_{00,01}=s, K_{00,10}=s,
 K_{00,11}=d$\\ $K_{01,00}=s, K_{01,01}=0, K_{01,10}=d, K_{01,11}=s$\\
 $K_{10,00}=s, K_{10,01}=d,  K_{10,10}=0, K_{10,11}=s$\\ $K_{11,00}=d,
 K_{11,01}=s,  K_{11,10}=s, K_{11,11}=0$\\  $s$ denotes  the  value of
 $K_{ab}$ when $a$ and $b$ differ  in a single replica and $d$ denotes
 its value  if $a$ and $b$ differ  in two replicas. Thus  a 16 element
 matrix only depends  on two parameters because of  the symmetries. In
 general,  the   $2^n  \times  2^n$  matrix  $K_{ab}$   has  just  $n$
 independent entries after taking the permutation symmetries into
 account. These  can be  parametrized by $K(p)$,  where $K(p)$  is the
 value of $K_{ab}$ when the states  $a$ and $b$ differ in $p$ replicas
 and  $p=1,2,  \cdots, n$.   There  is  a  similar simplification  for
 $Y_{ab}$ which  can be parametrized by  $Y(p)$ if the  states $a$ and
 $b$ differ in $p$ replicas.   In the given example of the $4\times
 4$ matrix, (the case of $n=2$), $s=K(1)$ and $d=K(2)$.

The  $2^n \times  2^n$ matrix  $K_{ab}$ is  readily  diagonalized \cite{Weigt,Coolen}.  In
general there  are $n+1$ distinct eigenvalues,  whose degeneracies are
$1,n,  n(n-1)/2, \cdots,\binom{n}{p},  \cdots,1$.  The  sum  of these
degeneracies is $2^n$ as required. There is a trivial eigenvalue equal
to $\sum_{p=1}^{n}\binom{n}{p}K(p)$ of  degeneracy unity.  If one sets
$K(p)\propto [(n-2p)-n]$, then there  is one other non-zero eigenvalue
of degeneracy $n$ and its eigenvectors identify it as corresponding to
a replicated  ferromagnetic Hamiltonian.   If $K(p) \propto  k(p)$, as
defined  in Eq.  (\ref{defkp}), there  is again  only  one non-trivial
non-vanishing eigenvalue,  of degeneracy $n(n-1)/2$,  which would be
natural for the spin glass state. Note that for ferromagnets the order
parameter  in  the  replicated  Hamiltonian is  $\langle  S_i^{\alpha}
\rangle$,  which  has  $n$  components,  while the  spin  glass  order
parameter,  $\langle S_i^{\alpha}S_i^{\beta}\rangle$,  has $n(n-1)/2$
independent components.

 The fixed  point studied by  KAS of Eqs. (\ref{RGY})  and (\ref{RGK})
 was  $K_{ab}=K^{*}$  and  $Y_{ab}=Y^*$   for  all  $a  \ne  b$.   The
 degeneracy of the non-trivial  eigenvalue of $K_{ab}^{*}$ is $2^n-1$.
 This fixed point therefore corresponds to a multicritical fixed point
 at which all the order parameters like $\langle S_i^{\alpha}\rangle$,
 $\langle       S_i^{\alpha}S_i^       {\beta}\rangle$,       $\langle
 S_i^{\alpha}S_i^{\beta}S_i^{\gamma}\rangle$  etc.,    go   critical
 simultaneously.   The critical exponents  associated with  this fixed
 point are those of the corresponding long-range $Q$-state Potts model
 \cite{AK}, with  $Q=2^n$, (where as usual  $n$ is set to  zero at the
 end  of the  calculation). It  is clearly  not the  appropriate fixed
 point for the study of the paramagnetic to spin glass transition.
 
 There  is a  further  additional  symmetry in  the  problem, that  of
 time-reversal symmetry \cite{AK}. Notice  that the bare kink coupling
 of Eq.  (\ref{defkp}) is left unchanged by the exchange $p\rightarrow
 n-p$.  This is  a manifestation of the invariance  of $K_{ab}$ in Eq.
 (\ref{Kdef})     under    the     time     reversal    transformation
 $\boldsymbol{\sigma}\rightarrow-\boldsymbol{\sigma}$.   This symmetry
 of  the  initial  couplings  is  again  maintained  by  the  RG  flow
 equations. Since  $K(0)=0$ by construction,  time reversal invariance
 implies  that  $K(n)=0$  and  in general  that  $K(p)=K(n-p)$.   Time
 reversal  symmetry  implies that  the  fugacities  obey the  relation
 $Y(p)=Y(n-p)$; in particular that  $Y(n)=0$ as $Y(0)=0$.  We could in
 principle reduce further the  number of independent kink coupling and
 their  associated  fugacities using  this  symmetry  but  it is  more
 convenient not to do this.

 Khurana \cite{AK} re-expressed the RG equations of Eqs. (\ref{RGY}) and
(\ref{RGK})  as  RG  equations  for   $K(p,l)$  and  the  associated
$Y(p,l)$.   He found  that  by  setting $K(p,l))=  K(l)  k(p)$, the  RG
equations for $K(p,l)$ reduce to just a single equation for $K(l)$:
\begin{equation} 
\frac{dK(l)}{dl}=\left[2(1-\sigma)-8\sum_{r=1}^{n-1}\binom{n-2}{r-1}Y(r,l)^2\right]K(l),
\label{rgK}
\end{equation}
with the  initial value  $K(l=0)=\beta^2J^2/2$. The RG  flow equations
for $Y(p,l)$ are more complicated:
\begin{eqnarray}
&{}&\frac{dY(p,l)}{dl}=Y(p,l)\left(1+\frac{2K(l)k(p)}{2\sigma-1}\right)+\label{rgY}\\
        &{}&\sum_{m=0}^{p}\binom{p}{m}\sum_{r=0}^{n-p}\binom{n-p}{r}Y(m+r,l)Y(r+p-m,l).
        \nonumber
\end{eqnarray}
Note  that $Y(0,l)=Y(n,l)\equiv  0$  and $p=1,  2,  \cdots, n-1$.  The
initial values are
\begin{equation}
Y(p,l=0)=\exp[U(0)K(l=0)k(p)],
\label{initial}
\end{equation}
where,     following     Cardy,     \cite{Cardy}     $U(0)     \approx
 1+\gamma+\beta^2J^2/2$.  $\gamma \approx 0.577$ is Euler's constant.

Khurana found a fixed point of these equations which is probably
  the  correct solution  for  $n>2$.   There is  a  fixed point  where
  $Y^*(1)=Y^*(n-1)$  is  of order  $(1-\sigma)^{1/2}$,  and the  other
  $Y^*(p)$ are  of order $(1-\sigma)$ or higher  powers.  The critical
  temperature for $\sigma \rightarrow 1$ was given by
\begin{equation}
 1+\frac{2K^*k(1)}{2\sigma-1} \approx 1+[(n-2)^2-n^2]\beta_c^2J^2=0.
\label{TcAK}
\end{equation}
The transition temperature  $T_c=2(n-1)^{1/2}J$ becomes pure
imaginary when $n  < 1$!  Hence this fixed  point of Eqs.  (\ref{rgK})
and (\ref{rgY})  cannot be  the appropriate fixed  point for  the spin
glass where we have to take $n=0$.

As they  stand the RG  equations do not  seem to make any  sense. They
hold  only for  small fugacities  $Y(p,l)$; their  derivation neglects
terms  of $O(Y^4)$  \cite{Cardy}.   But the  initial  values of  these
fugacities,  $\exp[U(0)\beta^2J^2k(p)/2]$ is  extremely  large in  the
low-temperature  limit where  $\beta$ goes  to infinity.   The  key to
progress is  to focus  on the sums  over fugacities in  equations like
(\ref{rgK}) and (\ref{rgY}).  By virtue of the limit $n\rightarrow
0$,  these individually  large terms  when summed  give rise  to small
well-behaved expressions. As a consequence we can find  solutions
of the KAS RG equations in the  regime in which they are expected to be accurate,
that is, for $\sigma$ approaching unity.  This is the subject of the next
section.

\section{Solving the RG flow equations}
\label{flows}
We start by solving Eqs. (\ref{rgY}) for $Y(p,l)$. We will make a guess that the solutions are of the form 
\begin{equation}
Y(p,l)=y(l)\exp[A(l)^2k(p)/8],
\label{Ansatz}
\end{equation}
when $p$ is $1,2, \cdots, n-1$, $y(0)=1$ and
\begin{equation}
\frac{A(l)^2}{8}=\frac{2}{2\sigma-1}\int_0^l\,dl^{\prime}K(l^{\prime})+
\frac{1}{2}\beta^2J^2U(0).
\label{Adef}
\end{equation}
Substituting this into Eq. (\ref{rgY}) gives after some algebra, which is  done in Appendix\ref{AppendixA},
\begin{equation}
\frac{dy(l)}{dl}=y(l)-4y^2(l).
\label{yeq}
\end{equation}
Corrections  to it are exponentially
 small, of $O(\exp[-1/(1-\sigma)^2])$  (see Appendix A) and so can  be dropped  in the limit $\sigma \rightarrow 1$. It  has solution
\begin{equation}
y(l)=\frac{1}{4-3e^{-l}},
\label{ysol}
\end{equation}
 showing that $y(l)$ decreases from its initial value of unity to its fixed point value
$y^*=1/4$.

We now turn to Eq. (\ref{rgK})  for the RG flow of $K(l)$. We shall  substitute the expression $Y(p,l)$ from Eq. (\ref{Ansatz}) into Eq. (\ref{rgK}), and set $n=0$ in $k(r)$ at the outset (keeping it in to the end and then setting it to zero gives the same result, but lengthens the equations). For $n=0$, $k(r)=4r^2$. We now need to evaluate the sum
\begin{equation}
S=\sum_{r=1}^{n-1}\binom{n-2}{r-1}\exp[A(l)^2r^2].
\label{Sdef}
\end{equation}
An integral representation allows $S$ to be re-written as
\begin{equation}
S=\int_{-\infty}^{\infty}\,\frac{dx}{\sqrt{2\pi}}\sum_{r=1}^{n-2}\binom{n-2}{r-1}\exp[\sqrt{2}A(l)xr-x^2/2]
\label{S1}
\end{equation}
which allows the sum over $r$ to be performed. Then 
\begin{equation}
S=\int_{-\infty}^{\infty}\,\frac{dx}{\sqrt{2\pi}}\exp[{\sqrt2}A(l)x-x^2/2](1+\exp[\sqrt{2}A(l)x])^{n-2}
\end{equation}
which becomes after setting once more $n$ to zero and some re-arrangement
\begin{equation}
S=\frac{1}{4}\int_{-\infty}^{\infty}\,\frac{dx}{\sqrt{2\pi}} \exp[-x^2/2]\mbox{sech}^2[A(l)x/\sqrt{2}].
\label{S3}
\end{equation}
We shall proceed upon the assumption that $A(l)$ is very large. Below we will show that its minimum value (see Eq. (\ref{Amin})) is of $O(1/(1-\sigma))$. For large $A(l)$,  in the integral over $x$ in Eq. (\ref{S3}), $\exp[-x^2/2]$ can be replaced by unity because the $\mbox{sech}$  term rapidly becomes very small for values of $x\sim1/A(l)$. The integral can now be evaluated directly giving 
\begin{equation}
S=\frac{1}{2\sqrt{\pi}A(l)}.
\end{equation}
Our final expression for the RG flow of $K(l)$ is
\begin{equation}
\frac{dK(l)}{dl}=\left[2(1-\sigma)-\frac{4y(l)^2}{\sqrt{\pi}A(l)}\right]K(l),
\label{rgKf}
\end{equation}
with the initial condition that $K(0)=\beta^2J^2/2$.

We have thus reduced the RG equations to an integro-differential equation. (The integral is in the definition of $A(l)$).  Eq. (\ref{rgKf})  could of course be
solved numerically. However, there is a simplification which allows for an
 analytical solution when $\sigma$ is close to unity. Eq. (\ref{ysol}) shows that $y(l)$ approaches its fixed point as a function of $e^{-l}$.  However, we shall see that
 $K(l)$ varies with $l$ much more slowly -- as a function not of $l$ but the combination $2(1-\sigma)l$. Since Eq. (\ref{rgKf}) is itself 
only valid for $\sigma$ close to unity, we can proceed by setting in it $y(l)$ to its fixed point value
$1/4$. This is because $y(l)$ will effectively be at its fixed point value before $K(l)$ has started to
 flow. Then Eq. (\ref{rgKf}) becomes analytically tractable. We shall also set 
$2\sigma-1$ which occurs in Eq.~(\ref{Adef}) to $1$ which is correct to leading order in
 $1-\sigma$.

Let $z(l)=A(l)^2/8$.  $z(0)=\beta^2J^2U(0)/2$. The first derivative 
\begin{equation}
\frac{dz(l)}{dl}\equiv\dot{z}=2K(l)
\label{zdot}
\end{equation}
and $\dot{z}(0)=\beta^2J^2$. Its second derivative 
\begin{equation}
\ddot{z}(l)=2 \frac{dK(l)}{dl}=2(1-\sigma)\dot{z}(l)-\frac{\dot{z}(l)}{\sqrt{128\pi z(l)}}.
\label{2deriv}
\end{equation}
 This equation can be integrated to 
\begin{equation}
\dot{z}(l)=2(1-\sigma)z(l)-\sqrt{z(l)/32\pi}+C,
\label{1int}
\end{equation}       
with the constant of integration
\begin{equation}
 C=\beta^2J^2-(1-\sigma)\beta^2J^2U(0)+\sqrt{\beta^2J^2U(0)/64\pi}.
\label{Cdef}
\end{equation}
Eq. (\ref{1int}) can be regarded as a quadratic equation for $\sqrt{z(l)}$, with solution
\begin{multline}
\sqrt{z(l)}=\frac{1\pm\sqrt{1-Q(C-2K(l))}}{\sqrt{512\pi}(1-\sigma)}\\
=\frac{1\pm\sqrt{(1-T_c/T)^2+2Q(K(l)-K(0))}}{\sqrt{512\pi}(1-\sigma)},
\label{zsol}
\end{multline}
where 
\begin{equation}
T_c = 16\sqrt{\pi U(0)}(1-\sigma)J,
\label{Tc}
\end{equation}
and $Q=256\pi(1-\sigma)$.
As a consequence of these manipulations we now have an ordinary differential equation for $K(l)$;
\begin{multline}
\frac{dK(l)}{dl}=2(1-\sigma)K(l) \times \\
\left[1-\frac{1}{1\pm \sqrt{(1-T_c/T)^2+2Q(K(l)-K(0))}}\right].
\label{Kfin}
\end{multline}
 Inspection of this equation shows that $K(l)$ will indeed be a function of $2(1-\sigma)l$, which justifies
 our setting in Eq. (\ref{rgKf}) $y(l)$ to its fixed point value $1/4$.

It may be more intuitive to express the flow equation for $K(l)$ as a flow equation for $T(l)$,
 the effective temperature on the lengthscale $ae^l$, defined by setting
\begin{equation}
K(l)=\frac{J^2}{2T(l)^2}.
\end{equation}
Then
 \begin{multline}
\frac{dT(l)}{dl}=-(1-\sigma)T(l) \times \\
\left[1-\frac{1}{1\pm \sqrt{(1-T_c/T)^2+QJ^2(1/T(l)^2-1/T^2)}}\right]
\label{Tfin}
\end{multline}
with the initial condition that $T(0)=T$.

Fixed points may arise if $dT(l)/dl=0$. For $T<T_c$ we  take the
 solution with the plus sign, while for $T>T_c$,
 we take the solution with the minus sign.
 There is a  fixed point at 
 $T(l)=0$, the zero-temperature fixed point. 
 The  \textit{unstable} critical fixed point is at $T=T_c$, $T(l)=T_c$
 If $T<T_c$, $T(l)$ flows from $T$  to the zero-temperature sink $T(l)=0$. When $T>T_c$, $T(l)$ increases with $l$, but eventually
it reaches a value at which the argument of the square root sign in Eq. (\ref{Tfin}) becomes zero and beyond that point, $T(l)$ goes complex. We conclude that our solution has no validity for $T>T_c$. From now on we will 
always take $T \le T_c$.

The derivation of the RG equation for $K(l)$ required
$A(l)$  to  be  large.  From Eq. (\ref{zsol}) the minimum value of $A(l)$ occurs at $l=0$ when $T=T_c$,
\begin{equation}
A(0)\rightarrow \frac{1}{8\sqrt{\pi}(1-\sigma)}.
\label{Amin}
\end{equation}
This is indeed large in the limit when $\sigma\rightarrow 1$ but  if $\sigma$ is fixed
 at some value close to unity, say $0.95$,  $A(l)$ will \textit{not} be large enough to make the approximation procedure reliable near and above $T_c$. However, we can work at a fixed value of $\sigma$ when  $T<<T_c$.  Then  $A(0)$ becomes very large:
 \begin{equation}
 A(0) \approx \frac{T_c}{8\sqrt{\pi}(1-\sigma)T},
 \label{ATsmall}
 \end{equation}
  and our solution will  become essentially exact.  Furthermore, as $A(l)$ increases to infinity as $l \rightarrow \infty$ when $T<T_c$, then our calculations of quantities like $\mathcal{S}(l)$, defined below, which are directly related to $A(l)$,  become essentially exact when $l$ is large.
  
Fortunately  some results can be obtained without needing an explicit  solution of the RG equations.  Eq.  (\ref{rgK}) can be re-written as a flow equation for $T(l)$,
\begin{equation}
\frac{dT(l)}{dl}=-T(l)[1-\sigma- \mathcal{S}(l)],
\label{rgsT}
\end{equation}
where the sum $\mathcal{S}(l)$ is defined to be
\begin{equation}
\mathcal{S}(l)= 4\sum_{r=1}^{n-1}\binom{n-2}{r-1}Y(r,l)^2.
\label{Sfull}
\end{equation}
Our calculation of the sum $\mathcal{S}(l)$ gave $\mathcal{S}(l)=2y(l)^2/[\sqrt{\pi}A(l)]$, which approaches zero as $l \rightarrow \infty$ when $T<T_c$. This large $l$ result should be exact even for fixed values of $\sigma$. It is fundamentally a consequence of the fact that
 at low temperatures there will be 
few large thermally excited droplets.  
Hence in the vicinity of the zero-temperature fixed point,
\begin{equation}
\frac{dT(l)}{dl} \rightarrow -(1-\sigma)T(l) =  -\theta T(l) \,\,\,\mbox{as}\,\,\, l \rightarrow \infty,
\label{theta}
\end{equation}
so the droplet  scaling exponent $\theta  =1-\sigma$, in
accordance with  conventional expectations\cite{KY1,KY2,FH,BMY}. In  fact one
expects this result  to hold for values of $\sigma$  not just close to
$1$ but throughout the range $2/3<\sigma<2$.

The transition  temperature $T_c$ can be determined  from Eq. (\ref{Tc})
by solving  the quadratic equation  which arises from  the temperature
dependence of $U(0)$. The result is
\begin{multline}
\left[\frac{T_c}{J}\right]^2=(1+\gamma)Q(1-\sigma)/2\\+\sqrt{(1+\gamma)^2Q^2(1-\sigma)^2+2Q(1-\sigma)}/2.
\label{Tcsol}
\end{multline}
As    $\sigma     \rightarrow     1$,    $T_c     \rightarrow
\sqrt{\sqrt{128\pi}(1-\sigma)}J$.   Thus  no finite temperature  transition  is  expected
right at $\sigma =1$.  This is  to be contrasted with what happens for
long-range Ising ferromagnets and  Potts models which \textit{do} have
a  finite  temperature  transition  at  their own lower critical value of $\sigma$\cite{Cardy}.  The
difference occurs because  at $T=0$ in an Ising  ferromagnet all spins
are aligned. No  kinks are present in the  low-temperature phase. They
appear  for $T>T_c$  when their  entropy overcomes  the energy  cost of
their creation.  In the leading term in $\overline{Z^n}$ as $T \rightarrow 0$, 
the replicas at
all sites have their spins  aligned\cite{AK}. The energy of this state
varies  as  $n^2$,  so does  not  contribute  to  the actual physical 
free  energy.  The leading term for the free energy contains flipped
spins and kinks\cite{AK}, and gives  a contribution to the free energy
of $O(n)$. The kink fugacities are always non-zero at all temperatures
in the spin glass.

 The   RG  equations   for  the   case   of  $\sigma=1$   are  as   in
 Eqs. (\ref{rgKf},  \ref{1int}) and (\ref{Cdef}), with  $\sigma$ there set
 to unity. The resulting flow equation for $K(l)$ is
\begin{equation}
\frac{dK(l)}{dl}=-\frac{K(l)}{64\pi[\beta
J\sqrt{U(0)/64\pi}+2(K(0)-K(l))]}.
\label{rgsig1}
\end{equation}
The  flow  starts  from  $K(0)=\beta^2J^2/2$  which  is  large  at  low
 temperatures.  We can  determine the  correlation length  $\xi(T)$ by
 finding  the scale  $l^*$ at  which $K(l)$  is of  order  unity. Then
 $\xi(T)=ae^{l^*}$. By integrating up  Eq. (\ref{rgsig1}) one obtains to
 leading order
\begin{equation}
\frac{\xi(T)}{a}\sim                        \exp[64\pi(\beta^2J^2+\beta
J\sqrt{U(0)/(64\pi)})\ln \beta^2J^2].
\end{equation}  
Thus as $T\rightarrow 0$, $\xi(T)$ diverges extremely rapidly.

\section{Behavior in a magnetic field}
\label{sec:field}
It is instructive to study the behavior of the system in a random magnetic field and in particular, the magnetic exponent $y_h$ using the RG formalism. 
This exponent describes the scaling dimension of the field conjugate to the spin-glass 
order parameter. Such a field is the Gaussian random field $h_i$ of zero mean and standard deviation $h$. Its presence  changes the Hamiltonian to 
\begin{equation}
H=-\sum_{<ij>}J_{ij}S_iS_j -\sum_ih_iS_i.
\label{Hamrf}
\end{equation}
After replicating and averaging over the the couplings $J_{ij}$ and the fields $h_i$, 
an additional term appears in the replicated Hamiltonian $\beta H_n$ of Eq. (\ref{RP}):
\begin{equation}
-\frac{1}{2}\beta^2h^2\sum_i\left(\sum_{\alpha=1}^{n}S_i^{\alpha}\right)^2.
\label{conjug}
\end{equation}
Let $\boldsymbol{t}$ denote the $n$-component vector $(1,1,1,\cdots,1)$. $\boldsymbol{t}$ can be used to re-write the magnetic field term in the replicated Hamiltonian as
\begin{equation}
-\frac{1}{2}N\beta^2h^2n-\sum_i\frac{1}{2}\beta^2h^2[(\boldsymbol{t} \cdot \boldsymbol{\sigma}_a(i))^2-n].
\label{ha}
\end{equation}
It is convenient to express the non-trivial term in Eq. \~ (\ref{ha}) in terms of $H_a(i)\equiv \mathcal{H}[(\boldsymbol{t} \cdot \sigma_a(i))^2-n]$, where $\mathcal{H}=\beta^2 h^2/2$.
$\mathcal{H}$ is the conjugate field for  
\begin{equation}
\frac{1}{N}\sum_i\sum_{\alpha \ne \beta} \langle S_i^{\alpha}S_i^{\beta}\rangle,
\end{equation}
the spin glass order parameter. The magnetic exponent $y_{\mathcal{H}}$ is
related  to the critical exponent $\eta$ via
\begin{equation}
y_{\mathcal{H}}=(d+2-\eta)/2,
\end{equation}
in dimension $d$.  With the long-range interactions  we are studying,
$2-\eta=2\sigma-1$\cite{Sak}, so $y_{\mathcal{H}}=\sigma$.  The question of interest is whether this result, which is valid for $2/3 <\sigma <1$, can be recovered from the KAS RG equations.

In a previous paper \cite{Moore2}  I derived the RG equations satisfied by the $H_a$ when they are small enough so that higher terms in $H_a$ can be neglected;
 \begin{equation}
\frac{dH_a(l)}{dl}=H_a(l) +\sum_bY_{ab}(l)^2[H_b(l)-H_a(l)],
\label{rgHa}
\end{equation}
for $a=1, \cdots, 2^n$. Remarkably these $2^n$ equations  reduce to a single equation for $\mathcal{H}(l)$ when $Y_{ab}(l)$ is expressed in terms of $Y(r,l)$:
\begin{eqnarray}
\frac{d\mathcal{H}(l)}{dl}&=&\left[1-4\sum_{r=1}^{n-1}\binom{n-2}{r-1}Y(r,l)^2\right]\mathcal{H}(l) \nonumber\\
&=&[1-\mathcal{S}(l)]\mathcal{H}(l)
\label{rgH}
\end{eqnarray}
This reduction to a single equation does not require any assumptions on the $r$ dependence of the $Y(r,l)$; it is a general result, of similar  origin to the reduction of the RG equations for $K_{ab}(l)$ to the single equation for $K(l)$ in Eq. (\ref{rgK}).

We first encountered the sum  $\mathcal{S}(l)$  in Eq. (\ref{rgsT}).  Its value at the critical fixed point is $1-\sigma$  when $l=0$. This will be true generically and not just for the limit  $\sigma \rightarrow 1$. Then  at $T=T_c$, the RG equation for $\mathcal{H}(l)$ reduces to 
\begin{equation}
\frac{d\mathcal{H}(l)}{dl}=\sigma \mathcal{H}(l)\,\,\,\mbox{as}\,\,\, l\rightarrow 0,
\end{equation}
implying that $y_{\mathcal{H}}=\sigma$, as anticipated.  One would expect this result to hold for $2/3<\sigma <1$.  It is reassuring that we can recover the exact value for the magnetic exponent $y_{\mathcal{H}}$ over this interval as our argument in Section \ref{sec:PQ} for replica symmetry in the same  range  proceeds along similar lines.

  As  we have noted before, $\mathcal{S}(l)$ approaches zero when $T<T_c$ as $l \rightarrow \infty$, when the RG flow is to the zero-temperature fixed point.  Hence
\begin{equation}
\frac{d\mathcal{H}(l)}{dl}=\mathcal{H}(l) \,\,\, \mbox{as} \,\,\,l\rightarrow \infty,
\end{equation}
which means that on length scale $L=a{\rm e}^l$,  because $\mathcal{H}=\beta^2h^2/2$, the effective field on a droplet of size $L$, $h(L)$, scales as $L^{d/2}$,with $d=1$,  as expected in Refs. [\onlinecite{McM, BM, FH}].  This result will hold for all $\sigma>2/3$.

\section{The Parisi overlap function $P(q)$}
\label{sec:PQ}
Whether the ordered phase of the spin glass is replica symmetric or not is determined by the form of the Parisi overlap function $P(q)$.  Parisi\cite{ParisiPq} showed that this could be calculated via
\begin{equation}
P_J(q)=\left \langle\delta\left(q-\frac{1}{N}\sum_iT_iS_i\right)\right \rangle,
\end{equation}
where the thermal average is over the doubly replicated Hamiltonian $H\{T_i\}+H\{S_i\}$. It is convenient to study 
\begin{equation}
F_J({\rm y})=\left \langle \left(\exp\frac{{\rm y}}{N}\sum_iT_iS_i\right)\right \rangle,
\end{equation}
as 
\begin{equation}
P_J(q)=\int_{-i\infty}^{i\infty}\frac{d{\rm y}}{2\pi i}e^{-{\rm y}q}F_J({\rm y}).
\label{Lap}
\end{equation}
The bond average, $\overline{F_J({\rm y})}(=F({\rm y}))$, leads to the determination of the Parisi overlap function;
$P(q)=\overline {P_J(q)}$. $F({\rm y})$ can be calculated by replicating the spins at each site, $S_i^{\alpha}, T_i^{\alpha}, \alpha=1, \cdots, n$.  The details of the calculation of $F({\rm y})$ in the context of the kink  RG procedure can be found in Ref. [\onlinecite{Moore2}]. The variable $\boldsymbol{\sigma}_{a}$ is now a vector of $2n$ components, where the first $n$ components contain the $\boldsymbol{S}$ spins while the second $n$ components are the $\boldsymbol{T}$ spins. $a$ now runs from $1$ to $2^{2n}$. The term in ${\rm y}$ produces in the replicated Hamiltonian $\beta H_{2n}$ a term  
\begin{equation}
-\sum_iH_a(i)=-\frac{{\rm y}}{N}\sum_i S^1_a(i)T^1_a(i).
\end{equation}
$S^1_a$ and $T^1_a$ are the first and $(n+1)$th components of $\boldsymbol{\sigma_a}$.  The RG flow equations for  the $H_a=({\rm y}/N)S^1_aT^1_a$ turn out to be identical in form to those in Eq. (\ref{rgHa})\cite{Moore2}, 
 \begin{equation}
\frac{dH_a(l)}{dl}=H_a(l) +\sum_bY_{ab}(l)^2[H_b(l)-H_a(l)],
\label{rgHaq}
\end{equation}
but now $a=1, \cdots, 2^{2n}$. There is another remarkable reduction of these $2^{2n}$ equations to a single equation when $Y_{ab}$ is expressed in terms of $Y(r,l)$.  The equations become the single equation for ${\rm y}(l)$,
\begin{equation}
\frac{d{\rm y}(l)}{dl}=\left[1-4\sum_{r=1}^{2n-1}\binom{2n-2}{r-1}Y(r,l)^2\right]{\rm y}(l).
\label{rgyq}
\end{equation}
The initial condition on this equation is that ${\rm y}(0)={\rm y}/N$.  Eq.  (\ref{rgyq}) is general;  it holds for any $Y(r,l)$. 

The sum in Eq. (\ref{rgyq}) is just the same as that which defines $\mathcal{S}(l)$ in Eq. (\ref{Sfull}), provided we replace $n$ in that equation by $2n$. In the limit of $n\rightarrow 0$ they will therefore be identical.  Integrating 
\begin{equation}
{\rm y}(l)=({\rm y}/N)\exp\left(l-\int_0^l dl^{\prime}\,\mathcal{S}(l^{\prime})\right).
\label{ysol2}
\end{equation}
At the value of $l=l^{*}$, where ${\rm e} ^{l^*}=N$, there are just two spins $S$ and $T$ left in the system; the other spins have been integrated out.
Their interaction is ${\rm y}(l^*)$. Then 
\begin{equation}
F({\rm y})=\frac{{\rm Tr}_{T,S}\exp[{\rm y}(l^*)TS]}{{\rm Tr}_T (1){\rm Tr}_S (1)}
=(e^{{\rm y}q_{EA}}+e^{-{\rm y}q_{EA}})/2.
\end{equation}
Here 
\begin{equation}
q_{EA}=\exp\left(-\int_0^{l^*} dl \,\mathcal{S}(l)\right), \,\,\,{\rm as}\,\,\, l^*\rightarrow\infty.
\label{qea1}
\end{equation}
Using Eq. (\ref{Lap}), one obtains
\begin{equation}
P(q)=[\delta(q-q_{EA})+\delta(q+q_{EA})]/2. 
\label{pqf}
\end{equation}
Eq. (\ref{pqf}) means that the spin glass phase has replica symmetry. It is the main result of this paper. The replica symmetric form arises from the reduction of Eq.  (\ref{rgHaq}) to a single equation. It does not require us to take the limit $\sigma \rightarrow 1$. We expect it to be valid for $2/3<\sigma<1$.

In the limit $\sigma \rightarrow 1$ we were able to show that $S(l)=2y(l)^2/[\sqrt{\pi}A(l)]$.  (Do not confuse ${\rm y}(l)$ with $y(l)$). On substituting for $y(l)$ its fixed point value of $1/4$
 and using Eqs. (\ref{ysol}) and (\ref{Kfin}) one obtains 
\begin{equation}
q_{EA}=\exp -\left (\frac{\pi/2-\mbox{tan}^{-1}(1/b)}{(T_c/T-1)b}\right),
\label{qea}
\end{equation}
where $b^2=QJ^2/(T_c-T)^2-1$. When $T \rightarrow 0$, $q_{EA} \rightarrow 1$, as expected. Less expected is the value of $q_{EA}$ as $T\rightarrow T_c^-$.  From Eq. (\ref{qea}), 
\begin{equation}
q_{EA}(T_c)= \exp[-(2\pi)^{3/4}/16].
\end{equation}
The conventional expection is that as $T \rightarrow T_c$, $q_{EA} \sim (1-T/T_c)^{\beta}$ so that $q_{EA}(T_c)=0$.   This is what will occur at a fixed value of $\sigma$. The finite value of $q_{EA}(T_c)$ might just be a peculiarity  of our $\sigma \rightarrow 1$  procedure.

\section{Discussion}
\label{sec:disc}

The chief conclusion of the paper is that the spin glass state of this
long-range one-dimensional Ising spin glass model is replica symmetric
when $2/3 <\sigma <1$.  This suggests that short-range $d$ dimensional
spin glasses will be replica symmetric when $d<6$.

I could only obtain an \textit{explicit} solution of the KAS RG equations for the limit
when $\sigma \rightarrow 1$ and $T<T_c$. The procedure for obtaining a solution of these equations for 
a fixed value of $\sigma$ close to unity, both above and below $T_c$ remains to be found. Fortunately the argument for replica symmetry does not depend on having an  explicit solution, but on general properties 
of the KAS equations.

In order to get the power law behavior which is expected at a fixed value of $\sigma$, the flow equation for $T(l)$ when it is close to $T_c$  will have to have  the linearized form
\begin{equation}
\frac{dT(l)}{dl}=\frac{1}{\nu}(T(l)-T_c),
\label{linT}
\end{equation}
where $\nu$ is the correlation length exponent. The flow will start from its initial value $T$. This equation can be used in  Eq.  (\ref{qea1}) when 
\begin{equation}
q_{EA}(T)=\exp\left(-\int_T^0\,dT(l) \frac{dl}{dT(l)}(1-\sigma)\right),
\end{equation}
where we have set $\mathcal{S}(l)$ to its fixed point value $1-\sigma$ and used the fact that when $T<T_c$,  $T(l)$  starts at $T$ and flows to zero. On evaluating the integral 
\begin{equation}
q_{EA}(T)=(1-T/T_c)^{\beta},  
\end{equation}
 where $\beta=\nu(1-\sigma)$.  This is just the result which follows from the scaling relation $\beta =\nu(d-2+\eta)/2=\nu(1-\sigma)$, when   $2-\eta=2\sigma-1$  \cite{Sak}. 
 
 The flow equation for $T(l)$ given in Eq. (\ref{Tfin}), which holds when $\sigma \rightarrow 1$, is not of the linearized form of Eq. (\ref{linT}) when $T(l)-T_c$ is small. However, as $l\rightarrow 0$, it does reduce to
 \begin{equation}
 \frac{dT(l)}{dl} \rightarrow (1-\sigma)(T-T_c)=(1-\sigma)(T(0)-T_c),
 \end{equation}
 It is therefore tempting to speculate that 
 \begin{equation}
 \nu=\frac{1}{1-\sigma},
 \end{equation}
 when $\sigma$ is close to unity.  
 
Alas, in order to check this speculation, the solution of the KAS RG equations at fixed $\sigma$  is required.

\begin{acknowledgments}
I should  like to  thank Helmut Katzgraber,  Derek Larson and Peter Young  for their communications
and insights  on  long-range one-dimensional systems over many years and Alan Bray for
   discussions over several decades.\\
\end{acknowledgments}

\appendix
\section{The RG equations for the fugacities} 
\label{AppendixA}
In this Appendix we indicate in more detail how the set of $n-1$ RG flow equations for the fugacities,  Eqs. (\ref{rgY}),  are reduced by means of  the guess in Eq.  (\ref{Ansatz}) to just a single simple RG equation, that in Eq. (\ref{yeq}), and explain why this is  essentially exact when $\sigma$ is close to $1$. 

We  proceed  by  substituting  the  guess  in  Eq.  (\ref{Ansatz})  into
Eq.  (\ref{rgY}).  According   to  the  guess  if  $p=0$   or  if  $p=n$
$Y(0,l)=Y(n,l)=y(l)$, but these terms at $p=0$ or $p=n$ are zero. Our
strategy for solving the equations is to define a function $\tilde{Y}(p,l) =Y(p,l)$ when $p=1,2, \cdots,n-1$ and $\tilde{Y}(0,l)=y(l)=\tilde{Y}(n,l)$. Then 
\begin{multline}
\frac{d\tilde{Y}(p,l)}{dl}=\tilde{Y}(p,l)\left(1+\frac{2K(l)k(p)}{2\sigma-1}\right)-4y(l)
\tilde{Y}(p,l)\\
        +\sum_{m=0}^{p}\binom{p}{m}\sum_{r=0}^{n-p}\binom{n-p}{r}\tilde{Y}(m+r,l)\tilde{Y}(r+p-m,l).
        \nonumber
\end{multline}
If the term involving the summations can be ignored, this equation has solution for 
$\tilde{Y}(p,l)$ as in Eq. (\ref{Ansatz}). Our task is thus to show that the summations produce only a very small correction. Define $R$ as
\begin{equation}
\sum_{m=0}^{p}\binom{p}{m}\sum_{r=0}^{n-p}\binom{n-p}{r}
\exp[A^2(k(m+r)+k(r+p-m))/8].
\nonumber
\end{equation}
Setting once again $k(p)=4p^2$, with the aid of a double integral representation
\begin{multline}
R=\int_{-\infty}^{\infty}\frac{dx}{\sqrt{2\pi}}\int_{-\infty}^{\infty}\frac{dy}{\sqrt{2\pi}}\sum_{m=0}^{p}\binom{p}{m}\sum_{r=0}^{n-p}\binom{n-p}{r} \\
\exp[A(m+r)x+A(r+p-m)y -(x^2+y^2)/2].\nonumber
\end{multline}
The sums over $r$ and $m$ can now be done explicitly when $R$ becomes (after setting $n$ to zero again) 
\begin{equation}
\int_{-\infty}^{\infty}\frac{dx}{\sqrt{2\pi}}\int_{-\infty}^{\infty}\frac{dy}{\sqrt{2\pi}}\exp[-(x^2+y^2)/2]\left(\frac{e^{Ax}+e^{Ay}}{1+e^{A(x+y)}}\right)^p.\nonumber
\end{equation}
The integrals can be simplified by introducing the new variables $u=(x+y)/2$, and $v=x-y$. Then $R=R_uR_v$ where
\begin{equation}
R_u=\int_{-\infty}^{\infty}\frac{du}{\sqrt{2\pi}}e^{-u^2}\mbox{sech}^p(Au), 
\nonumber
\end{equation}
and 
\begin{equation}
R_v=\int_{-\infty}^{\infty}\frac{dv}{\sqrt{2\pi}}e^{-v^2/4}\cosh^p(Av/2).
\nonumber
\end{equation}
When $A(l)$ is large, as it always at temperatures $T \le T_c$, the integals for $R_s$ and $R_d$ can be done by steepest descents. $R_u$ is of $O(1/A(l))$ while
 $R_v \sim \exp[A(l)^2k(p)/16]$. This means that the corrections to Eq. (\ref{yeq}) from $R$ are
 of order $\exp[-A(l)^2k(p)/16]$.  The smallest value of $A(l)$  is of order $1/(1-\sigma)$ according to Eq. (\ref{Amin}), so provided that $\sigma$ is close to $1$, these corrections should be negligible.


\begin{thebibliography}{99}

\bibitem{Leuzzi:99} L. Leuzzi,  J. Phys. A {\bf 32}, 1417 (1999).

\bibitem{KY1} H. G. Katzgraber and A. P. Young,  Phys. Rev. B {\bf 67},
134410 (2003).

\bibitem{KY2} H. G. Katzgraber and A. P. Young,  Phys. Rev. B {\bf 68},
224408 (2003).

\bibitem{Leuzzi:08} L.   Leuzzi, G.  Parisi,  F.  Ricci-Tersenghi, and
 J. J. Ruiz-Lorenzo,  Phys. Rev. Lett.  {\bf 101}, 107203  (2008).

\bibitem{Hart08}   H.    G.    Katzgraber   and  A.    K.    Hartmann,
Phys. Rev. Lett., {\bf 102},  037207  (2009).

\bibitem{Larson} H.   G.  Katzgraber, D.   Larson, and A.   P.  Young,
Phys. Rev. Lett., {\bf 102}, 177205 (2009).

\bibitem{KAS}  G.   Kotliar,  P.   W.   Anderson,  and  D.  L.   Stein,
Phys. Rev.  B {\bf 27}, R602  (1983).

\bibitem{Larsonpspin} D. Larson, H. G. Katzgraber, M. A. Moore, and A. P. Young,
Phys. Rev. B {\bf 81}, 064415 (2010). 

\bibitem{Moore1} M.  A. Moore,  J. Phys.  A: Math. Gen.  {\bf 38}, L783
  (2005).
  
\bibitem{Parisi}  G.  Parisi,  Phys.   Lett.  {\bf  73A}, 203  (1979);
J. Phys. A: Math. Gen. {\bf 13}, L115 (1980).

\bibitem{McM} W. L. McMillan,  J. Phys. C {\bf 17}, 3179 (1984).

\bibitem{BM} A.  J.  Bray and M. A.  Moore,  Lecture Notes in Physics,
{\bf 275}, 121 (1987).

\bibitem{FH} D. S.  Fisher and D. A. Huse,  Phys. Rev.  B {\bf 38}, 386
  (1988).

\bibitem{Leuzzi2009} L. Leuzzi, G. Parisi, R. Ricci-Tersinghi, and J. J. Ruiz-Lorenzo,
Phys. Rev. Lett., {\bf 103}, 267201 (2009). 

\bibitem{Cardy} J.   L. Cardy,  J. Phys.  A:  Math. Gen.  {\bf 14}, 1407
  (1981).

\bibitem{Moore2} M.  A. Moore, J. Phys.  A:  Math. Gen.  {\bf 19}, L211
  (1986).
  
\bibitem{AK} A. Khurana,  Phys. Rev. B {\bf 40}, 2602 (1989).

\bibitem{Weigt} M. Weigt and R. Monasson, Europhys. Lett. {\bf 36}, 209 (1996).

\bibitem{Coolen} T. Nikoletopoulos and A. C. C. Coolen,
 J. Phys. A: Math. Gen. {\bf 37}, 8433 (2004).
 
 \bibitem{BMY}  A.   J.   Bray,  M.   A.  Moore,  and  A.   P.   Young,
Phys. Rev. Lett. {\bf 56}, 2641 (1986).

\bibitem{Sak} J. Sak, Phys. Rev. B {\bf 8},  281 (1973).

\bibitem{ParisiPq} G. Parisi, Phys. Rev. Lett. {\bf 50},  1946 (1983).




\end{thebibliography}
\end{document}